\documentclass[aps,prapplied,twocolumn,superscriptaddress]{revtex4-2}

\usepackage{graphicx} 
\usepackage{amsmath} 
\usepackage{mhchem} 
\usepackage[colorlinks,linkcolor=blue,citecolor=green,urlcolor=blue]{hyperref} 
\usepackage{color} 

\begin{document}


\title{Quantitative Assessment of Carrier Density by Cathodoluminescence. I. GaAs thin films and modeling}


\author{Hung-Ling Chen}
\author{Andrea Scaccabarozzi}
\affiliation{Centre de Nanosciences et de Nanotechnologies (C2N), CNRS, Universit\'e Paris-Saclay, 91120 Palaiseau, France}

\author{Romaric De L\'epinau}
\affiliation{Centre de Nanosciences et de Nanotechnologies (C2N), CNRS, Universit\'e Paris-Saclay, 91120 Palaiseau, France}
\affiliation{Institut Photovolta\"ique d'Ile-de-France (IPVF), 91120 Palaiseau, France}

\author{Fabrice Oehler}
\author{Aristide Lema\^itre}
\author{Jean-Christophe Harmand}
\author{Andrea Cattoni}
\affiliation{Centre de Nanosciences et de Nanotechnologies (C2N), CNRS, Universit\'e Paris-Saclay, 91120 Palaiseau, France}

\author{St\'ephane Collin}
\affiliation{Centre de Nanosciences et de Nanotechnologies (C2N), CNRS, Universit\'e Paris-Saclay, 91120 Palaiseau, France}
\affiliation{Institut Photovolta\"ique d'Ile-de-France (IPVF), 91120 Palaiseau, France}
\email[]{stephane.collin@c2n.upsaclay.fr}

\date{\today}

\begin{abstract}

Doping is a fundamental property of semiconductors and constitutes the basis of modern microelectronic and optoelectronic devices. Their miniaturization requires contactless characterization of doping with nanometer scale resolution. Here, we use low- and room-temperature cathodoluminescence (CL) measurements to analyze p-type and n-type GaAs thin films over a wide range of carrier densities ($2\times 10^{17}$ to $1\times 10^{19}$~cm$^{-3}$). The spectral shift and broadening of CL spectra induced by shallow dopant states and band filling are the signature ofƒ doping. We fit the whole spectral lineshapes with the generalized Planck law and refined absorption models to extract the bandgap narrowing (BGN) and the band tail for both doping types, and the electron Fermi level for n doping. This work provides a rigorous method for the quantitative assessment of p-type and n-type carrier density using CL. Taking advantage of the high spatial resolution of CL, it can be used to map the doping in GaAs nanostructures, and it could be extended to other semiconductor materials.

\end{abstract}


\maketitle


\section{Introduction}

Optimization of doping is one of the main requirements for functional semiconductor devices. Carrier concentrations in thin films are typically determined using capacitance-voltage profiling~\cite{Blood:1986} or Hall effect measurements~\cite{Schroder:2006}. Alternative contactless and non-destructive methods are desirable to obtain fast insights in the material quality and doping information in the early stage of semiconductor manufacturing. Raman spectroscopy provides signatures of doped polar semiconductors due to longitudinal optical (LO) phonon-plasmon interactions~\cite{Abstreiter:1984}. Photoluminescence (PL) is sensitive to the residual of dopants and PL spectra can change considerably in shape at high carrier concentrations~\cite{Pavesi:1994}. Redshifts of the PL spectra in p-type doping have been investigated~\cite{Casey:1976,Olego:1980a}, and the analysis of the bandgap narrowing (BGN) effect has been proposed as a nondestructive measurement of hole densities~\cite{Lu:1994}. On the other hand, the blueshift of the PL spectra, known as the Burstein-Moss effect~\cite{Burstein:1954,Moss:1954}, prevails in n-type III-V semiconductors. The band filling, the band tail and the BGN need to be considered to properly describe the luminescence spectra~\cite{Bugajski1985,Lee:1995,Chen:2017}. However, properties such as the BGN as a function of carrier concentrations reported in the literature deviate substantially, and an easy-to-use theoretical framework for the luminescence analysis of both p-type and n-type carrier concentrations is still lacking. Moreover, optical methods do not provide the spatial resolution required for the development of devices based on semiconductor nanostructures like nanowires or nanopillars~\cite{Goktas:2018,Eaton:2016}, or to probe polycrystalline thin films at the scale of single grains and grain boundaries~\cite{Powalla:2018}.

Cathodoluminescence (CL) can be used as an alternative method to assess the carrier density of n-type and p-type semiconductor layers with a spatial resolution as low as a few tens of nanometers. In a previous work, we have used high-resolution CL mapping to determine the doping of single nanowires made of n-type gallium arsenide (GaAs) with electron concentrations of about $10^{18}$~cm$^{-3}$~\cite{Chen:2017}. Here, we extend our previous study to p-type doping, we clarify the formalism and fitting procedures, and we use a larger number of reference samples to refine the determination of doping using luminescence spectra. We use reference, planar semiconductor layers to settle experimental methods and theoretical models for the assessment of both n-type and p-type doping over a wide range of concentrations. We focus our study on GaAs due to its widespread use in microelectronics such as microwave integrated circuits, high-frequency and low-noise amplifiers. Moreover, GaAs has a direct bandgap, allowing to fabricate efficient optoelectronic devices such as infrared lasers, light-emitting diodes, and solar cells~\cite{Schnitzer:1993a,Kayes:2011}.

In the following, we first describe the p-doped and n-doped GaAs thin films used in this work and we estimate the injection level of CL measurements. We present low-temperature CL measurements and we discuss qualitatively the effect of doping on the luminescence spectra. Then, room-temperature CL measurements are presented. The generalized Planck law and an absorption model for doped GaAs are used to fit the luminescence spectra. The important parameters fitted with this model (bandgap, Fermi level, Urbach absorption tail) are related to the carrier concentration. Our results are compared to the literature and to empirical models. This systematic, thorough investigation is general and can be adapted to other semiconductors. CL is particularly useful to inspect doping variations at the nanometer scale.

\section{\label{sec:exp}Experiments}


\begin{table*}
	\caption{\label{tab:samples}MBE-grown GaAs thin-film samples. Mobility and carrier density are obtained from Hall effect measurements.}
	\begin{ruledtabular}
	\begin{tabular}{lccclccc}
		\multicolumn{4}{c}{p-type GaAs:Be} & \multicolumn{4}{c}{n-type GaAs:Si} \\
		\cline{1-4} \cline{5-8}
		sample & thickness & mobility & carrier density & sample & thickness & mobility & carrier density \\
		& $t$ (nm) & $\mu_p$ (cm$^2$/Vs) & $p$ (cm$^{-3}$) & & $t$ (nm) & $\mu_n$ (cm$^2$/Vs) & $n$ (cm$^{-3}$) \\
		\cline{1-4} \cline{5-8}
		P1 & 500 & 243 & $2.7\times 10^{17}$ & N1 & 500 & 3220 & $2.2\times 10^{17}$ \\ 
		P2 & 500 & 153 & $9.4\times 10^{17}$ & N2 & 465 & 4066 & $4.0\times 10^{17}$ \\
		P3 & 500 & 135 & $2.4\times 10^{18}$ & N3 & 320 & 2750 & $9.4\times 10^{17}$ \\
		P4 & 380 & 122 & $2.9\times 10^{18}$ & N4 & 473 & 1980 & $1.8\times 10^{18}$ \\
		P5 & 430 & 74 & $1.0\times 10^{19}$ & N5 & 852 & 1850 & $3.9\times 10^{18}$ \\
		P6 & 500 & 78 & $1.4\times 10^{19}$ & N6 & 446 & 1165 & $7.6\times 10^{18}$ \\
	\end{tabular}
	\end{ruledtabular}
\end{table*}

GaAs thin films were grown by solid-source molecular beam epitaxy (MBE) on semi-insulating GaAs(001) substrates. \ce{As4} was supplied and a Ga flux was set to obtain the GaAs growth rate of about 0.2~nm/s. Beryllium (Be) and silicon (Si) are used for p-type and n-type doping of GaAs, respectively. Table~\ref{tab:samples} lists the set of GaAs thin-film samples. The total thicknesses of doped GaAs are measured from the cross-section scanning electron microscope (SEM) images of cleaved samples (typical uncertainty $\pm$10~nm).

Carrier mobility and concentration were determined by Hall effect measurements using the van der Pauw method~\cite{vanderPauw:1958}. No additional surface passivation layer was grown on top of the doped GaAs layers. The thickness of the electrically active region used to calculate the bulk carrier concentration was corrected for the surface depletion width~\cite{Look:1990}. Several samples listed in Table~\ref{tab:samples} were used in previous works with slightly different carrier densities~\cite{Chen:2017,Chen2018b}. During the preparation of this article, we repeated Hall effect measurements and analyses of all samples on two different setups and found inaccuracies for a series of samples grown and measured several years ago. The data provided here should replace the carrier densities published in references~\cite{Chen:2017,Chen2018b}, and can be used confidently as references.

CL measurements were performed using an Attolight Chronos cathodoluminescence microscope. Samples were excited by a focused electron beam with an acceleration voltage of 6~kV and an impinging current of about 0.7~nA. Luminescence is collected through an achromatic reflective objective (numerical aperture 0.72), dispersed with a Horiba diffraction grating (150 grooves/mm) and recorded with an Andor Newton CCD camera (1024$\times$256 pixels, pixel width 26~$\mu$m). The corresponding spectral dispersion in the visible to near infrared range is 0.53~nm per pixel~\cite{Chen:2017}. Luminescence spectra were corrected for the diffraction efficiency of the grating and the sensitivity of the CCD camera. 

It is important to ensure that the injection density does not impact the shape of luminescence spectra and the assessment of carrier densities~\cite{Mendis2019}. The injection regime of CL measurements can be estimated as follows. The total electron-hole pair generation rate $G$ is given by~\cite{Yacobi:1986}:
\begin{equation}
	G=\frac{(1-b)VI\eta}{qE_g},
\end{equation}
where $b$ is the back-scattering coefficient (flat GaAs $b\approx 30\%$), $V$ is the electron beam voltage and $I$ is the beam current, $q$ is the elementary charge and $E_g$ is the semiconductor bandgap (eV), $\eta$ is the quantum efficiency for carrier generation by the electron beam: $\eta\approx 30\%$ for typical semiconductors~\cite{Davidson:1980}. The interaction volume of the electron beam in matter can be simulated using the Monte Carlo method~\cite{Drouin:2007}. For a more practical approach, the energy loss distribution is approximately Gaussian~\cite{Davidson:1980}:
\begin{equation}
	g=\frac{4G}{\pi^{3/2}a^3}\text{exp}\left(-\frac{r^2}{a^2}\right)\text{cos}\:\theta,
\end{equation}
where $r$ is the distance to the excitation point in the semiconductor and $\theta$ is the angle with respect to the electron beam. To estimate the maximum excess carrier density $\Delta N$, the density of generation rate should be put into the diffusion equation, which is solved with appropriate boundary conditions. Ref.~\cite{Davidson:1980} gives the approximate solutions for infinite and zero surface recombination velocity $S$. In our case, we take $a=100$~nm for 6~kV electron beam in GaAs, a carrier lifetime $\tau=1$~ns and a diffusion length $L=1~\mu$m. With these values, $\Delta N=G\tau/(16L^2a)\approx 2.4\times10^{15}$~cm$^{-3}$ for $S=\infty$ and $\Delta N=G\tau/(5L^2a)\approx 8\times10^{15}$~cm$^{-3}$ for $S=0$. Therefore, the CL injection level should be much lower than the carrier concentrations studied here. Indeed, no spectral shift or shape variations of the main emission peak were observed in CL spectra when varying the excitation current from about 0.3 to 3~nA (see the appendix). Surface depletion is another effect that can impact the carrier concentration determined by CL. In the range studied here, the surface depletion width is much smaller than both the GaAs layer thickness and the e-beam penetration depth, and it is not expected to affect the shape of luminescence spectra~\cite{Chen:2019a}. On each sample, we have performed CL mapping over roughly $10 \times 10~\mu$m$^2$ and obtained homogeneous luminescence maps with typical peak energy variation within $\pm$2~meV. Throughout this contribution, we present CL spectra averaged over the CL map for each sample, and we analyze the effect of doping on CL spectra.

\section{\label{sec:lt_cl}Low-temperature CL}

\subsection{p-type GaAs}

\begin{figure}
	\includegraphics[width=\columnwidth]{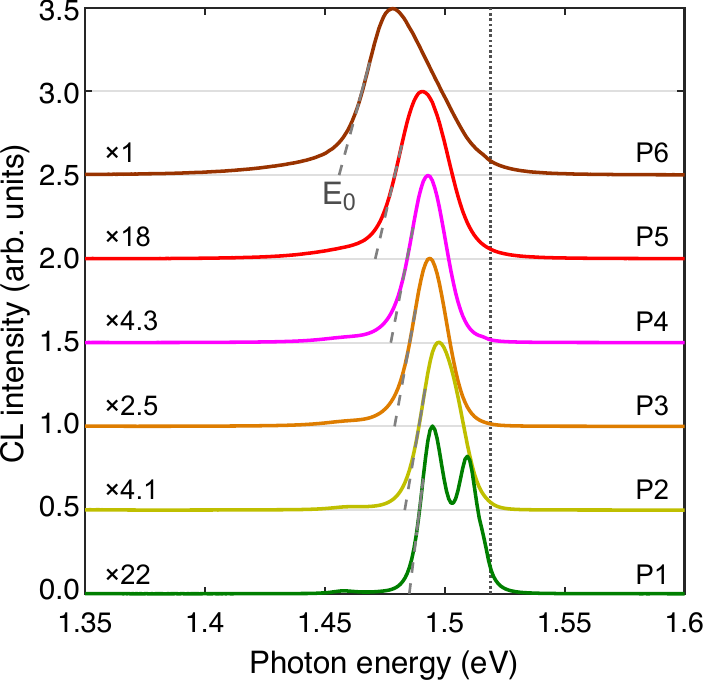}
	\caption{\label{fig:ptype_LT}CL spectra of p-GaAs:Be thin films measured at low temperature (20~K). CL intensities are normalized by their maximal intensity and shifted vertically for clarity. The $E_0$ bandgap is obtained by a linear extrapolation of the low-energy side of the spectrum to the background. The vertical dashed line indicates the bandgap of undoped GaAs (1.519~eV).}
\end{figure}

Fig.~\ref{fig:ptype_LT} shows the CL spectra of p-GaAs:Be samples measured at low temperature (LT, 20~K). The spectra are normalized to their maximum intensity and shifted vertically to facilitate their comparison. For the lowest doping (sample P1, green curve in Fig.~\ref{fig:ptype_LT}), the CL spectrum exhibits two separate peaks: 1.509~eV for the exciton bound to acceptor, and 1.495~eV for the transition of free electrons to Be acceptors. For higher doping levels, the acceptor band merges with the valence band, thus only one single emission peak is observed. The density corresponding to this transition is approximately $8\times 10^{17}$~cm$^{-3}$ for p-GaAs, calculated using the model of Ref.~\cite{Serre:1983}. Above this threshold, the emission spectra redshift with increasing doping concentrations due to the effect of BGN that is theoretically explained by carrier-carrier exchange and correlation and carrier-dopant ion interactions~\cite{Mahan:1980}. To quantify the effect of BGN, a method widely used is to assign a bandgap $E_0$ defined by the intersection between the tangent to the low-energy tail of the spectrum and the background~\cite{Olego:1980a} (gray dashed lines in Fig.~\ref{fig:ptype_LT}). The difference between $E_0$ and the bandgap of undoped GaAs (1.519~eV) defines a BGN value (see Section~\ref{sec:rt_cl_p} for further discussion).

\begin{figure}
	\includegraphics[width=\columnwidth]{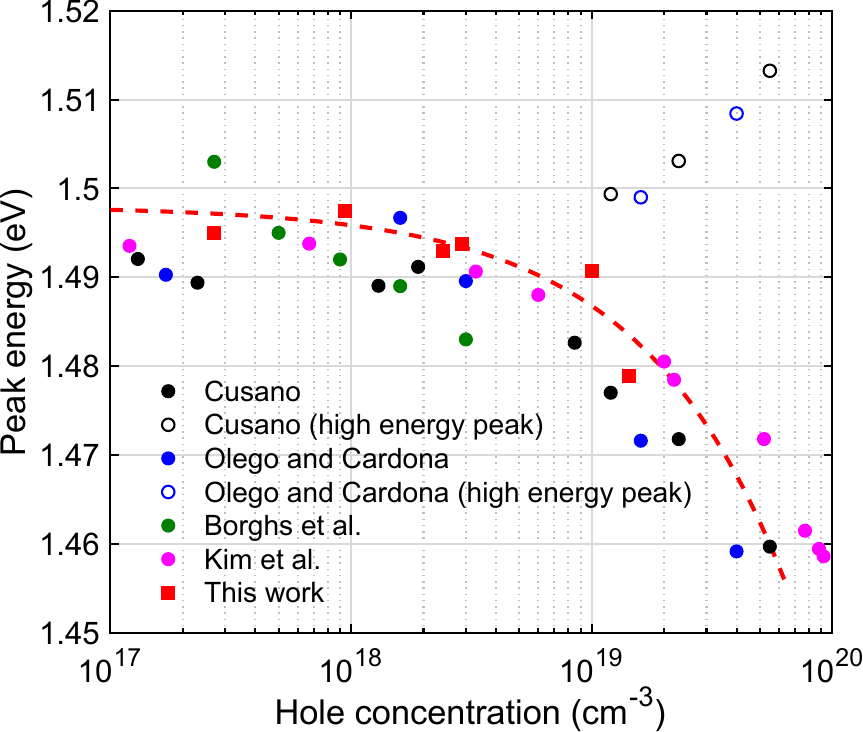}
	\caption{\label{fig:hole_peak}Peak position energy as a function of the hole concentration. Our CL spectra are compared to available literature data from Cusano for 20~K CL of GaAs single crystal slices~\cite{Cusano:1965}, Olego and Cardona for 4~K PL of Zn-doped single crystal GaAs~\cite{Olego:1980a}, Borghs et al. for 30~K PL of MBE-grown Be-doped GaAs~\cite{Borghs:1989} and Kim et al. for 12~K PL of MOCVD-grown C-doped GaAs~\cite{Kim:1993a}. The red dashed line is a guide to the eyes.}
\end{figure}

In Fig.~\ref{fig:hole_peak}, we compare our results with data taken from the literature. The PL or CL peak position measured at LT are plotted as a function of the hole concentration for various p-GaAs doping. Our CL results were fitted with three Gaussian functions in order to capture the eventual asymmetrical shape and extract the peak position in an accurate manner (red squares in Fig.~\ref{fig:hole_peak}). We can see the trend of BGN with increasing doping concentrations. The spread of data points may be due to different sample preparations, uncertainties related to carrier concentration measurements or luminescence analysis. Open circles shows the high energy peak observed in PL or CL spectra in degenerate p-GaAs (carrier density above $\sim 10^{19}~\text{cm}^{-3}$). This feature is referred to as Mahan exciton and is caused by the absorption singularity at the Fermi level~\cite{Mahan:1967}. We did not observe a clear high energy peak in our highest doped sample (P6).

\subsection{n-type GaAs}

\begin{figure}
	\includegraphics[width=\columnwidth]{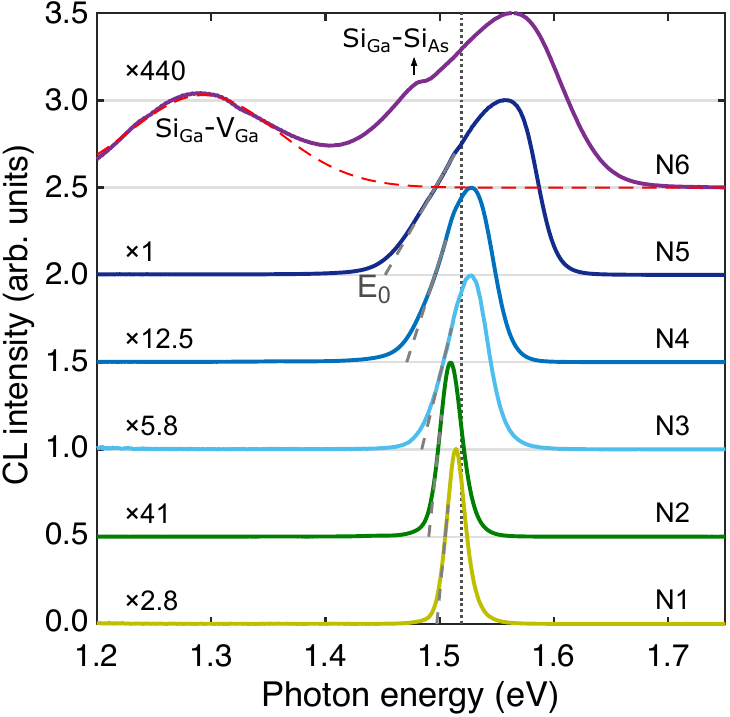}
	\caption{\label{fig:ntype_LT}CL spectra of n-GaAs:Si thin films measured at low temperature (20~K). CL intensities are normalized by their maximal intensity and shifted vertically for clarity. Except for sample N6, the $E_0$ bandgap is obtained by a linear extrapolation of the low-energy side of the spectrum to the background. The vertical dashed line indicates the bandgap of undoped GaAs (1.519~eV).}
\end{figure}

Fig.~\ref{fig:ntype_LT} shows the CL spectra of n-GaAs:Si samples measured at LT (20~K). For lightly doped samples (N1 and N2), the CL peak is mainly due to the shallow donor band (typical donor ionization energy: 6~meV). With increasing n-doping levels, the CL spectra gradually broaden and the peak positions shift to higher energies.
It is observed predominantly in n-type III-V semiconductors because the conduction band filling is more common owing to the relatively small effective density of conduction band states ($N_c\approx 4.2\times 10^{17}~\text{cm}^{-3}$ for GaAs). We note that the BGN effect is also present in n-GaAs and results in the shift of the low-energy tail of CL spectra towards lower energies with increasing doping levels.

For the highest n-doped sample (N6, purple curve in Fig.~\ref{fig:ntype_LT}), a shoulder near 1.48~eV may be attributed to the recombination involving Si acceptor (\ce{Si_{As}})~\cite{Pavesi:1994}. Furthermore, a wide Gaussian-like signal at around 1.29~eV is visible. This broad emission band is commonly observed in highly Si-doped GaAs, and may be attributed to the recombination involving the complex of Si donor (\ce{Si_{Ga}}) and Ga vacancy (\ce{V_{Ga}}, deep acceptor)~\cite{Pavesi:1992,Ky:1998}. Theses features have been observed in Si-doped GaAs layers under surface thermal annealing~\cite{Chiang:1975} and in compensated GaAs:Si samples grown using liquid-phase epitaxy~\cite{Kressel:1968,Kressel:1969}. 

\begin{figure}
	\includegraphics[width=\columnwidth]{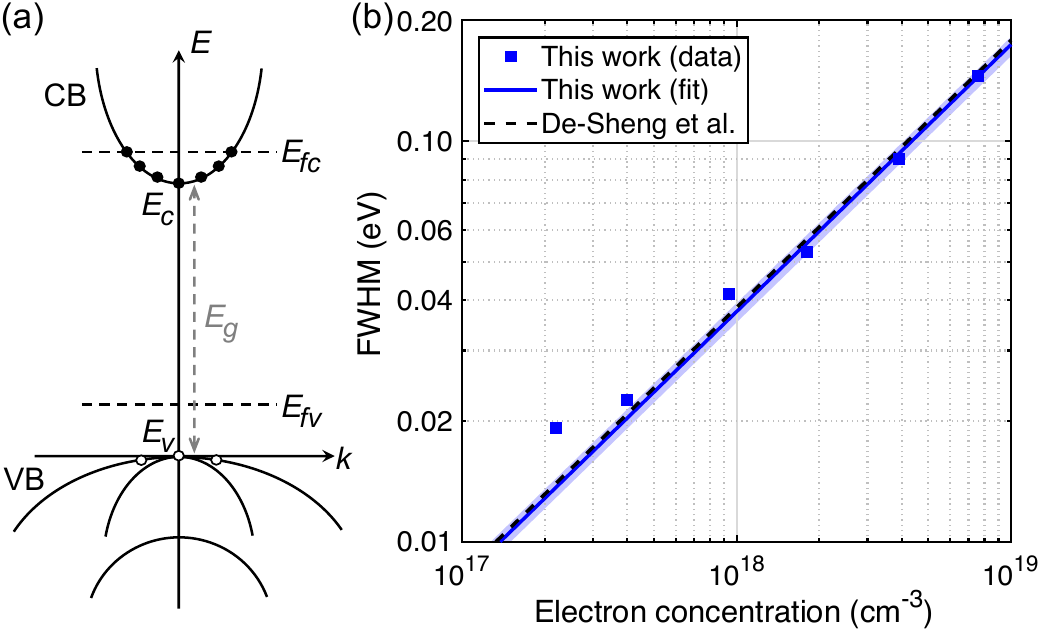}
	\caption{\label{fig:electron_width}(a) Schematic band diagram of a degenerate n-type III-V semiconductor. Under excitation, excess carriers are characterized by two separate quasi-Fermi levels. (b) FWHM measured at low temperature as a function of the electron concentration $n$. Black dashed line is from De-Sheng et al. \cite{De-Sheng:1982} for 1.8~K PL of MBE-grown n-GaAs:Te samples. Blue markers are CL measurements, the blue line and shaded area show the fit with a $n^{2/3}$ power law and the confidence interval, respectively (Eq.~\ref{eq:electron_fwhm}).}
\end{figure}

The electron concentration $n$ can be related to the full-width at half maximum (FWHM) of luminescence spectra measured at LT when thermal broadening is practically suppressed. A phenomenological formula was first established for Te-doped GaAs by De-Sheng et al.~\cite{De-Sheng:1982}. Fig.~\ref{fig:electron_width}(a) shows a schematic of the radiative recombination process in a degenerate n-type semiconductor. The Fermi level $E_{fc}$ is above the conduction band minimum and free electrons fill the states below $E_{fc}$. In a first-order approximation, the width of the luminescence spectrum scales proportionally to $E_{fc}$ (using $E_c$ as the zero reference) assuming that the spread in energy of holes is much smaller than that of electrons. As a 3D electron gas occupies a so-called Fermi sphere of radius $k_F=(3\pi n)^{1/3}$, $E_{fc}$ is proportional to $k_F^2$ (parabolic conduction band) and is thus proportional to $n^{2/3}$. Indeed, our experimental FWHM values can be fitted easily with a 2/3 power function of $n$:
\begin{equation}
	\text{FWHM~(eV)}=(3.75\pm0.21)\times 10^{-14}\times n^{2/3},
	\label{eq:electron_fwhm}
\end{equation}
where $n$ is the electron concentration expressed in $\text{cm}^{-3}$, and $\pm0.21$ indicates the $95\%$ confidence interval of the fitted parameter defined as $\pm t \sigma$, with $t$ calculated using the $t$-distribution, and $\sigma$ the standard error of the parameter. As shown in Fig.~\ref{fig:electron_width}(b), our result is very close to that of De-Sheng et al.~\cite{De-Sheng:1982}. This empirical relation between LT FWHM and the electron concentration $n$ (Eq.~(\ref{eq:electron_fwhm})) can be used to determine the electron concentration in the range of about $4\times 10^{17}$ to $1\times 10^{19}~\text{cm}^{-3}$.

\section{\label{sec:rt_cl}Room temperature CL and modeling}

We now focus on CL spectra measured at room temperature (RT), the usual operating condition of semiconductor devices. Defects and exciton recombination tend to dominate luminescence at LT, while excitons are practically dissociated and shallow donors/acceptors are totally ionized at RT. Band-to-band recombination of excess carriers is the most important mechanism at RT and luminescence spectra can be modeled more easily.

\begin{figure}
	\includegraphics[width=\columnwidth]{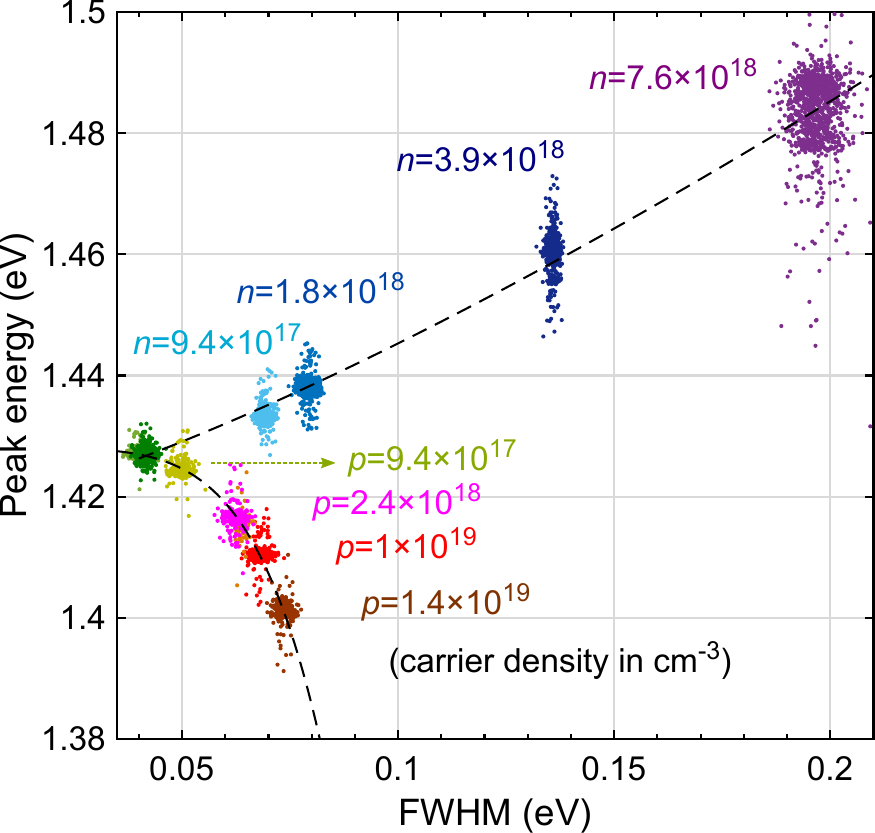}
	\caption{\label{fig:scattering_plot}Scattering plot of peak energy versus FWHM from room-temperature CL measurements on different n-doped and p-doped planar GaAs layers. Each dot represents the characteristics of a spectrum recorded as a pixel on CL maps. CL spectra broaden for increasing doping concentration, and the peak position shifts to higher energies for n-doping and to lower energies for p-doping. Dashed lines are guides to the eye.}
\end{figure}

Figure~\ref{fig:scattering_plot} shows a scattering plot of the peak energy versus the FWHM of CL spectra measured at RT. Each dot is defined by the characteristics of a spectrum extracted from a single pixel of CL hyperspectral maps. The FWHM continuously enlarges with increasing doping levels. 
However, similarly to the low-temperature case (Fig.~\ref{fig:ptype_LT}), the FWHM of p-type GaAs varies more slowly with the doping as compared to n-type GaAs, and can hardly be used to assess the carrier concentration. Figure~\ref{fig:scattering_plot} evidences the redshift of the peak energy induced by increasing p-type doping, while n-type doping results in an opposite blueshift of the peak. In the following, a more precise analysis of the luminescence lineshape is performed using the generalized Planck law together with an absorption model for doped GaAs. It will enable a quantitative determination of carrier densities.

\subsection{\label{sec:planck}Generalized Planck law and absorption model}

It has been known for a long time that the spectral distribution of radiative recombination is related to the absorption properties of a semiconductor~\cite{VanRoosbroeck1954}, and can be expressed explicitly to account for the band filling~\cite{Lasher1964}. Following the formalism of reference \cite{Wurfel:1982}, the luminescence spectrum $\phi(\hbar\omega)$, where $\hbar\omega$ is the photon energy, is expressed as a function of the absorptivity $A(\hbar\omega)$ and the quasi-Fermi levels of electrons ($E_{fc}$) and holes ($E_{fv}$) in a semiconductor:
\begin{equation}
	\phi(\hbar\omega) = \frac{A(\hbar\omega)}{4\pi^2\hbar^3c^2}\frac{(\hbar\omega)^2}{\text{exp}\left(\frac{\hbar\omega-(E_{fc}-E_{fv})}{k_BT}\right)-1},
	\label{eq:planck1}
\end{equation}
where $\hbar$ is the reduced Planck constant, $c$ the speed of light in vacuum, $k_B$ the Boltzmann constant and $T$ the absolute temperature.
It is called the generalized Planck law, and it was verified experimentally on a GaAs luminescent diode~\cite{Feuerbacher1990}. It is valid for semiconductors under quasi-thermal equilibrium, provided the quasi-Fermi levels can be defined. If the photon energy is several $k_BT$ larger than the separation of the quasi-Fermi levels, the $-1$ term in the denominator of Eq.~(\ref{eq:planck1}) can be neglected, leading to:
\begin{equation}
	\phi(\hbar\omega) \simeq A(\hbar\omega)\phi_\text{bb}(\hbar\omega)\text{exp}\left(\frac{E_{fc}-E_{fv}}{k_BT}\right),
	\label{eq:planck2}
\end{equation}
where $\phi_\text{bb} = (\hbar\omega)^2/4\pi^2\hbar^3c^2 \left[ \text{exp}\left(\hbar\omega/k_BT\right)-1 \right]^{-1}$ is the black-body radiation.

Doping has a significant impact on the absorption, especially in the spectral range near the bandgap. First, we denote $\alpha_\text{ideal}(\hbar\omega)$ the absorption coefficient of an ideally pure semiconductor. It can be calculated theoretically based on the joint density of states and the optical transition matrix element (e.g. the parabolic approximation for direct bandgap semiconductors), or measured on a high-purity semiconductor. With increasing doping, sub-bandgap absorption occurs and can be modeled by an exponential decay (Urbach tail) instead of an abrupt decrease of absorption for $\hbar\omega<E_g$:
\begin{equation}
	\alpha_0(\hbar\omega) \sim \text{exp}\left(-\frac{E_g-\hbar\omega}{\gamma}\right),
\end{equation}
where $\gamma$ is a parameter that characterizes the energy width of the Urbach tail. We use a convolution to join the Urbach tail with the ideal absorption~\cite{Katahara:2014}:
\begin{equation}
	\alpha_0(\hbar\omega) = \frac{1}{2\gamma} \int_{E_g}^{\infty} \alpha_\text{ideal}(\mathcal{E}) \text{exp}\left(-\frac{\left|\hbar\omega-\mathcal{E}\right|}{\gamma}\right)d\mathcal{E}.
	\label{eq:convolution}
\end{equation}

Second, the band filling effect needs to be included. The occupation probability is related to the Fermi functions $f_c$ and $f_v$ characterized by the quasi-Fermi levels for electrons and holes, respectively. The corrected absorption term is written as~\cite{Katahara:2014,Bhattacharya2012}:
\begin{subequations}
	\begin{eqnarray}
	& \alpha(\hbar\omega) = \alpha_0(\hbar\omega)\times \left(f_v-f_c\right), \\
	\label{eq:occupation1}
	& f_v - f_c = \frac{1}{\text{exp}\left(\frac{\mathcal{E}_h-E_{fv}}{k_BT}\right)+1} - \frac{1}{\mathrm{exp}\left(\frac{\mathcal{E}_e-E_{fc}}{k_BT}\right)+1},
	\label{eq:occupation2}
	\end{eqnarray}
\end{subequations}
where the photon energy is equal to the energy difference between electrons and holes: $\hbar\omega=\mathcal{E}_e-\mathcal{E}_h$. The excess energy $\hbar\omega-E_g$ is weighted between electrons and holes according to their respective effective masses using a parameter $w$ ($0<w<1$):
\begin{subequations}
	\begin{eqnarray}
	\mathcal{E}_e-E_c &=& w(\hbar\omega-E_g), \\
	\label{eq:parameter_w1}
	E_v-\mathcal{E}_h &=& (1-w)(\hbar\omega-E_g).
	\label{eq:parameter_w2}
	\end{eqnarray}
\end{subequations}
For most zinc-blende III-V semiconductors, $w$ close to 1 is a realistic approximation since the excess energy is taken by electrons rather than holes. For GaAs, with the electron effective mass of 0.063$m_0$ and the heavy hole effective mass of 0.50$m_0$, the ratio leads to $w=0.89$. We note that, at low injection levels, the occupation factor in Eq.~(\ref{eq:occupation2}) depends essentially on majority carriers (doping). The spread of the Fermi level for minority carriers, that may be induced by an inhomogeneous carrier density, will not influence the spectral lineshape of luminescence. For instance, in n-type semiconductors, $E_{fv}-E_v\gg k_BT$ and $f_v\sim 1$. According to Eq.~(\ref{eq:planck2}), the hole Fermi level only affects the total intensity through the exponential factor, not the shape of luminescence spectra. This enables the lineshape analysis to yield information on the doping, regardless the concentration of minority carriers.

Finally, the luminescence spectrum is described in term of absorptivity, which depends not only on the material's bulk property but also on its structure. In inhomogeneous or nanostructured materials, it can be described by a local absorption cross section~\cite{Greffet:2018}. For a homogeneously excited slab of thickness $d$, the absorptivity is given by~\cite{Wurfel:1982}:
\begin{equation}
	A(\hbar\omega) = \left(1-R)\left[1-\text{exp}\left(-\alpha(\hbar\omega\right) d\right)\right].
\end{equation}
$R$ is the reflectivity on the front surface, which should have minor impact on the luminescence lineshape as compared to the absorption drop near the bandgap. In general, $d$ can be regarded as a characteristic length scale over which carriers are generated, travel and recombine radiatively~\cite{Katahara:2014}. For $d\to 0$, the luminescence spectral shape is approximated by replacing $A(\hbar\omega)$ with $\alpha(\hbar\omega) d$ in Eq.~(\ref{eq:planck1}), which means the luminescence is produced near the top surface and reabsorption can be neglected. Otherwise, in the spectral region where $\alpha(\hbar\omega)d\gtrsim 1$ (usually in the high-energy tail of luminescence spectra), the effect of reabsorption may distort the luminescence lineshape. 

\subsection{\label{sec:rt_cl_p}Bandgap narrowing in p-GaAs}

\begin{figure*}
	\includegraphics[width=0.8\textwidth]{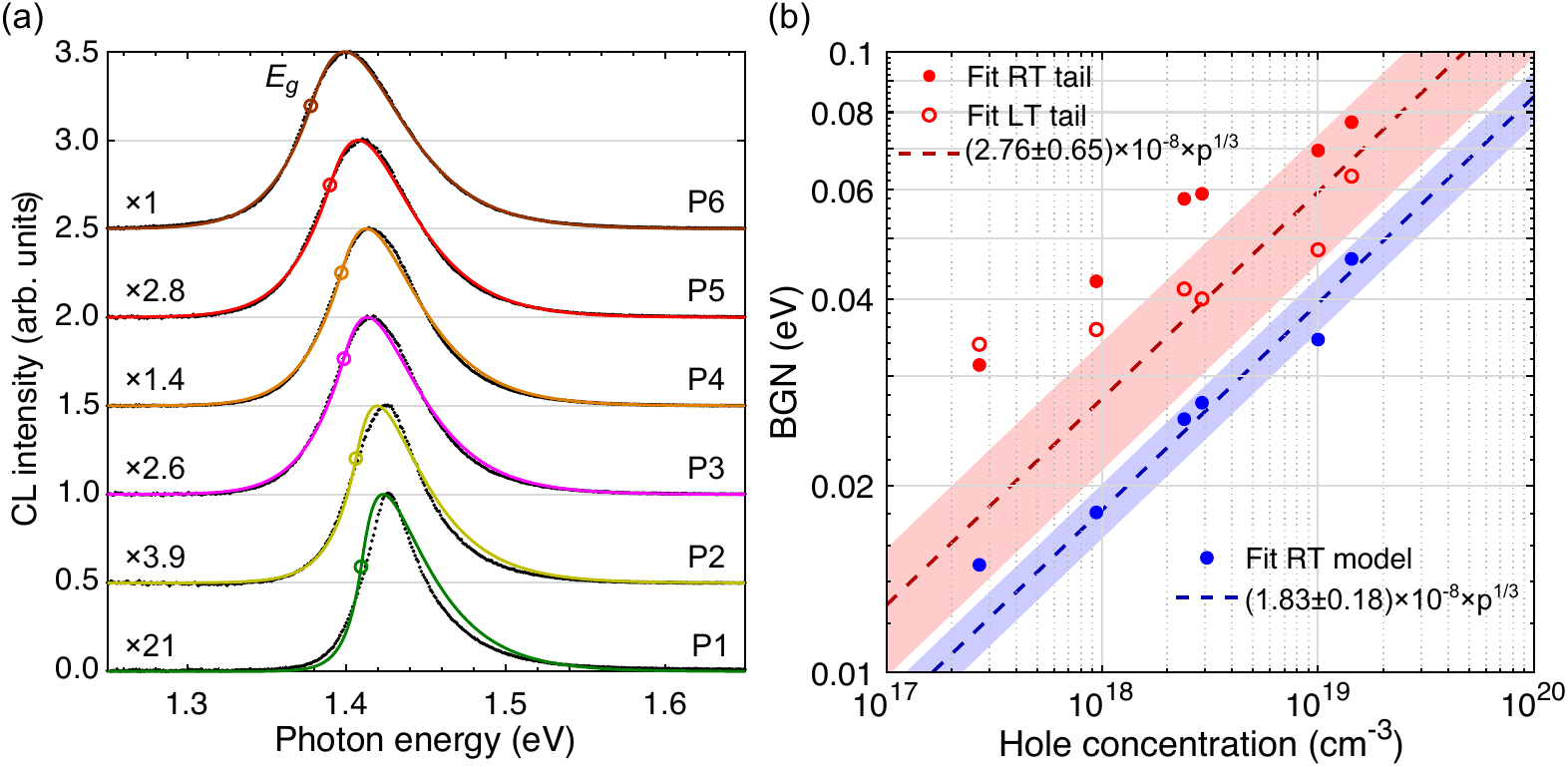}
	\caption{\label{fig:ptype_RT}(a) CL spectra of p-GaAs:Be thin films measured at RT (black dots) and the fitted generalized Planck law (colored lines). The CL normalization factors are indicated. Open circles superimposed on the CL spectra mark the bandgaps determined from the fits. (b) BGN as a function of hole concentration $p$. Blue dots are bandgaps obtained with a parabolic absorption model combined with Urbach tails (RT model),
and the blue solid curve is the fit given in Eq.~(\ref{eq:bgn_c2n}). Red dots and circles are bandgaps obtained from the linear extrapolation of the low-energy tail of RT and LT CL spectra, and the red line shows the fit given in 
Eq.~(\ref{eq:bgn_hlc}). The shaded areas show the confidence interval of the fits.}
\end{figure*}

Figure~\ref{fig:ptype_RT}(a) shows the RT CL spectra measured on p-GaAs:Be samples. They are fitted with the generalized Planck law using the least-squares method to determine the bandgap $E_g$ and Urbach tail $\gamma$. Since the absorption coefficient of doped GaAs at energies above 1.6~eV is rather independent of the doping level~\cite{Casey:1975}, we take a parabolic band for the ideal absorption coefficient with the value of $14800~\text{cm}^{-1}$ at $\hbar\omega=1.6$~eV for all doping~\cite{Casey:1975}: $\alpha_\text{ideal}=14800\sqrt{(\hbar\omega-E_g)/(1.6-E_g)}$ if $\hbar\omega>E_g$ and 0 otherwise. $\alpha_\text{ideal}$ is then convoluted with an Urbach tail following the description of Section~\ref{sec:planck}. The temperature $T$ is fixed at 300~K and $d$ is the thickness of the doped layer. The results of fits are plotted in Fig.~\ref{fig:ptype_RT}(a) (colored lines) and fitted parameters are given in Table~\ref{tab:pdoping_fit}. For low doping (P1 and P2), CL spectra differ from the fits due to excitonic enhancement of absorption near the bandgap. Better fits can be obtained if $d$ is varied to compensate for the low absorption in a simple parabolic model (see the appendix).

\begin{table}
	\caption{Peak energy, FWHM and optimal fit parameters (bandgap $E_g$ and Urbach tail $\gamma$) for CL spectra measured on p-GaAs thin films at room temperature. Carrier densities from Hall effect measurements are indicated.}
	\label{tab:pdoping_fit}
	\begin{tabular}{lcccccc}
		\hline\hline
		sample & carrier density & peak & FWHM & $E_g$ & $\gamma$ \\
		& cm$^{-3}$ & eV & eV & eV & meV \\
		\hline
		P1 & $2.7\times 10^{17}$ & 1.427 & 0.040 & 1.409 & 7  \\ 
		P2 & $9.4\times 10^{17}$ & 1.425 & 0.050 & 1.406 & 10 \\ 
		P3 & $2.4\times 10^{18}$ & 1.417 & 0.062 & 1.398 & 14 \\ 
		P4 & $2.9\times 10^{18}$ & 1.416 & 0.064 & 1.397 & 14 \\ 
		P5 & $1.0\times 10^{19}$ & 1.410 & 0.068 & 1.390 & 17 \\ 
		P6 & $1.4\times 10^{19}$ & 1.401 & 0.074 & 1.378 & 18 \\ 
		\hline\hline
	\end{tabular}
\end{table}

In Fig.~\ref{fig:ptype_RT}(b), we plot the BGN values as a function of hole concentration, together with empirical BGN from the literature. We show the BGN determined from different methods. Blue dots correspond to $E_g$ fitted using the generalized Planck law (RT model). For heavily doped p-GaAs, the bandgap narrowing is expected to vary with the hole concentration as $p^{1/3}$~\cite{Casey:1976}. Casey and Stern determined the bandgap by fitting the absorption measurements for three p-GaAs thin films with hole concentrations from $1.2\times 10^{18}$ to $1.6\times 10^{19}~\text{cm}^{-3}$ and found~\cite{Casey:1976}:
\begin{equation}
	E_g~(\text{RT}) = 1.424 - 1.6\times 10^{-8}\times p^{1/3},
	\label{eq:bgn_casey}
\end{equation}
where $p$ is the hole concentration in cm$^{-3}$ and $E_g$ in eV. Using six samples with hole concentrations spanning over a larger range, our fit should provide a slightly more accurate dependence of the bandgap narrowing:
\begin{equation}
	E_g~(\text{RT}) = 1.424 - (1.83\pm0.18)\times 10^{-8}\times p^{1/3}.
	\label{eq:bgn_c2n}
\end{equation}
Figure~\ref{fig:ptype_RT}(b) also includes red dots and circles, which correspond to the $E_0$ bandgap defined as the linear extrapolation of the low-energy tail to the background using RT and LT CL spectra, respectively. This method includes the contribution of band tails, and thus yields larger BGN values. Red dashed line is the empirical fit from Lu et al.~\cite{Lu:1994} using LT $E_0$ bandgaps for p-GaAs with holes concentrations varying between $3\times 10^{17}$ to $1\times 10^{20}~\text{cm}^{-3}$:
\begin{equation}
	E_0~(\text{LT}) = 1.519 - 2.4\times 10^{-8}\times p^{1/3},
	\label{eq:bgn_lu}
\end{equation}	
which has the same functional form as Eq.~(\ref{eq:bgn_casey}) but with a larger prefactor. For highly-doped samples (hole concentration above $10^{18}$~cm$^{-3}$), our BGN values from LT $E_0$ bandgap lie in the region of Eq.~(\ref{eq:bgn_lu}) and is in agreement with early experimental works~\cite{Borghs:1989,Kim:1993a} and many-body calculations~\cite{Sernelius:1986a,Jain:1990}. The fit of our dataset results in:
\begin{equation}
	E_0~(\text{LT}) = 1.519 - (2.76\pm0.65)\times 10^{-8}\times p^{1/3}.
	\label{eq:bgn_hlc}
\end{equation}
RT BGN values are $\sim$20~meV larger than LT BGN values. This temperature dependence was also observed by Lu et al. and is explained by the strong hole-phonon interactions at RT, resulting in a modification of the $E_g(T)$ relation at high hole densities~\cite{Lu:1994}.

In summary, the bandgap appears as a reliable quantity to determine the doping concentration of p-GaAs for $p>10^{18}$~cm$^{-3}$. RT luminescence spectra can be fitted to extract $E_g$ bandgap decoupled from the band tail, and then the BGN is related to $p$ through Eq.~(\ref{eq:bgn_casey}). Alternatively, LT $E_0$ gap is commonly used and Eq.~(\ref{eq:bgn_lu}) is adequate to determine the hole concentration.

\subsection{Band filling in n-GaAs}

\begin{figure*}
	\includegraphics[width=\textwidth]{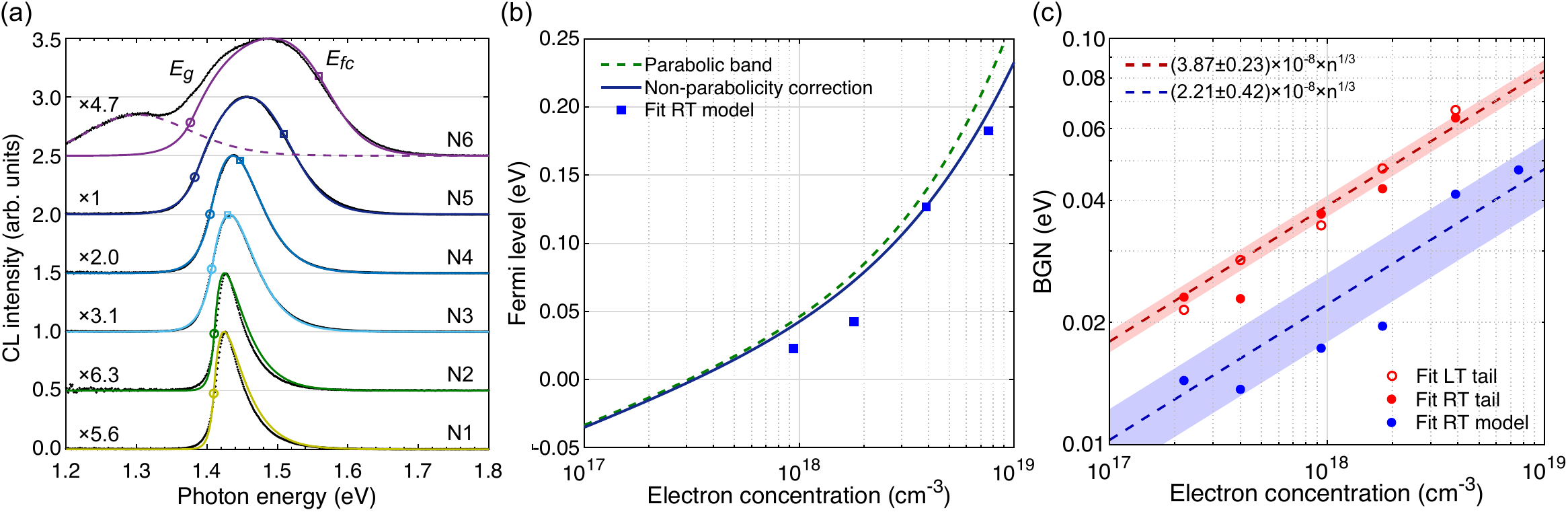}
	\caption{\label{fig:ntype_RT}(a) CL spectra of n-GaAs:Si thin films measured at RT (dots) and the fits (lines). The CL normalization factors are indicated. Circles and squares mark the fitted bandgaps and electron Fermi levels, respectively. (b) Electron Fermi level obtained from the fit of CL spectra (circles) and theoretical curves calculated by assuming a parabolic conduction band (dashed line) and non-parabolicity correction (solid line). (c) BGN as a function of electron concentration $n$ and fits of the datasets, see Eqs.~(\ref{eq:nEgRT}) and (\ref{eq:nE0}). The shaded areas show the confidence interval of the fits. Blue dots are obtained from the fit of CL spectra. Red dots and red circles are obtained from the linear extrapolation of low-energy tail of RT and LT CL spectra, respectively.}
\end{figure*}

CL spectra measured at RT on n-GaAs:Si samples are plotted in Fig.~\ref{fig:ntype_RT}(a). They are fitted using the same procedure as described previously, with $E_{fc}$ an additional fitting parameter and $E_{fv}$ fixed several $k_BT$ above the valence band edge. Fitted parameters are given in Table~\ref{tab:ndoping_fit} and results are shown in Fig.~\ref{fig:ntype_RT}(a): colored curves are the fitted model, circles and squares on each CL spectrum indicate the bandgaps and electron Fermi levels extracted from the fits, respectively. For low-doped samples (N1 and N2: $n<5\times 10^{17}$~cm$^{-3}$), accurate electron Fermi levels cannot be easily extracted from the luminescence analysis. For the mostly doped sample (N6), luminescence from deep levels is deconvoluted with a Gaussian term. Additional fits using an effective thickness are provided in the appendix.

\begin{table}
	\caption{\label{tab:ndoping_fit}Peak energy, FWHM and optimal fit parameters (bandgap $E_g$, electron Fermi level $E_{fc}$, Urbach tail $\gamma$) for CL spectra measured on planar n-GaAs at room temperature. Carrier densities from Hall effect measurements are indicated.}
	\begin{tabular}{lcccccc}
		\hline\hline
		sample & carrier density & peak & FWHM & $E_g$ & $\gamma$ & $E_{fc}$ \\
		& cm$^{-3}$ & eV & eV & eV & meV & meV \\
		\hline
		N1 & $2.2\times10^{17}$ & 1.427 & 0.042 & 1.410 & 5 & --   \\ 
		N2 & $4.0\times10^{17}$ & 1.428 & 0.042 & 1.410 & 5 & --   \\ 
		N3 & $9.4\times10^{17}$ & 1.433 & 0.069 & 1.407 & 13 & 23  \\ 
		N4 & $1.8\times10^{18}$ & 1.438 & 0.079 & 1.404 & 16 & 43  \\ 
		N5 & $3.9\times10^{18}$ & 1.461 & 0.136 & 1.383 & 20 & 127 \\ 
		N6 & $7.6\times10^{18}$ & 1.484 & 0.197 & 1.377 & 23 & 182 \\ 
		\hline\hline
	\end{tabular}
\end{table}

In Fig.~\ref{fig:ntype_RT}(b), the electron Fermi level $E_{fc}$ deduced from the fit of CL spectra are plotted as a function of the electron concentration. The dashed curve represents a theoretical relation between the electron concentration and the Fermi energy in GaAs using the effective density of the conduction band: $N_c=4.2\times 10^{17}~\text{cm}^{-3}$. The solid line represents the relation corrected for the non-parabolicity of the conduction band:
\begin{equation}
	n = N_c \left[\mathcal{F}_{1/2}\left(\frac{E_{fc}-E_c}{k_BT}\right)-\frac{15\beta k_B T}{4E_g}\mathcal{F}_{3/2}\left(\frac{E_{fc}-E_c}{k_BT}\right)\right],
	\label{eq:nonparabolicity}
\end{equation}
where
\begin{equation}
	\mathcal{F}_j(x) = \frac{1}{\Gamma (j+1)} \int_{0}^{\infty} \frac{t^j}{\mathrm{exp}(t-x)+1} dt
\end{equation}
is the Fermi integral of order $j$, $\Gamma$ is the gamma function, and the non-parabolicity factor $\beta\approx -0.83$ for n-GaAs at RT~\cite{Blakemore:1982}. The non-parabolicity of the conduction band results in a lower electron Fermi level at a given electron concentration. We obtained slightly lower $E_{fc}$ values than expected (solid line) for intermediately doped samples (N3 and N4). This may be due to surface depletion in n-GaAs~\cite{Look:1990}, non negligible electron occupation in the conduction band tail states~\cite{Lee:1995}, or reabsorption effect~\cite{Sieg:1996}. Still, this method should provide a quantitative determination of doping at high concentration ($1\times 10^{18}$ to $1\times 10^{19}~\text{cm}^{-3}$).

In Fig.~\ref{fig:ntype_RT}(c), we compare the BGN values of n-GaAs using different analysis methods. Blue dots correspond to $E_g$ extracted from the fit of RT CL spectra. Their trend with electron concentration $n$ can be fitted with (blue dashed line):
\begin{equation}
	E_g~(\text{eV}) = 1.424 - (2.21\pm0.42)\times 10^{-8} \times n^{1/3}.
\label{eq:nEgRT}
\end{equation}
Red dots (resp. circles) correspond to $E_0$ obtained by extrapolation of RT (resp. LT) low-energy tail of CL spectra. No significant temperature dependence of BGN values is observed for n-GaAs. These BGN values depend on the electron concentration as $n^{1/3}$:
\begin{equation}
	\text{BGN}~(\text{eV}) = (3.87\pm0.23)\times 10^{-8} \times n^{1/3}.
\label{eq:nE0}
\end{equation}
In the literature, BGN values of n-GaAs are very contentious. Early calculations predicted extremely large BGN for n-GaAs~\cite{Sernelius:1986,Jain:1990}. Borghs et al. reported BGN of n-GaAs nearly twice our values (green dashed line in Fig.~\ref{fig:ntype_RT}(c))~\cite{Borghs:1989}, probably due to presence of other impurities in their samples such that the luminescence spectra extend deeply into the bandgap. Our results are in close agreement with recent experimental works~\cite{Hudait:1999,Luo:2002}.

\section{Conclusion}

In conclusion, we have presented CL measurements on p-type and n-type GaAs thin-film samples over a large range of carrier densities, at low- and room-temperature. We have used the generalized Planck law together with refined absorption models including Urbach tails and band filling to fit the whole spectra and to extract the bandgap, band tail, and Fermi level.

For p-GaAs with hole concentration below approximately $8\times 10^{17}~\text{cm}^{-3}$, we observed two distinct CL peaks at LT due to recombination involving shallow acceptors. At higher hole concentrations, the acceptor band merges with the valence band thus only a single CL peak was observed. CL spectra continuously broaden and redshift with increasing carrier concentrations due to the BGN effect. The bandgaps $E_g$ were extracted by fitting RT CL spectra. Using Eq.~(\ref{eq:bgn_c2n}), they provide an accurate way to determine the carrier concentration of p-GaAs in the range of about $1\times 10^{18}$ to $2\times 10^{19}$~cm$^{-3}$.

For n-GaAs, the CL spectra steadily broaden and blueshift with increasing electron concentrations. The broadening at the low-energy side of the CL spectra is due to increased BGN and band tail at high doping levels, while the broadening at high-energies is the result of the conduction band filling. The bandgaps $E_g$ and electron Fermi levels were extracted by fitting RT CL spectra. The Fermi levels are useful to access the electron concentration in a range of about $1\times 10^{18}$ to $1\times 10^{19}$~cm$^{-3}$. The FWHM of LT CL spectra provides another accurate way to determine the electron concentration from about $4\times 10^{17}$ to $1\times 10^{19}~\text{cm}^{-3}$ (Eq.~(\ref{eq:electron_fwhm})).

Considering an uncertainty of the measured CL characteristics of about 2~meV or less (e.g., LT FWHM of n-GaAs and BGN of p-GaAs), and the good agreement between measurements and Eqs.~(\ref{eq:electron_fwhm}) and (\ref{eq:bgn_c2n}), the accuracy of the method is estimated at about $10\%$ to $20\%$, in the same order than typical Hall effect measurements used as references.

Using high-resolution CL mapping, we have applied the experimental methods and reference models developed in this paper for the determination of numerous n-type and p-type carrier densities of single GaAs nanowires. It will be the subject of a forthcoming article~\cite{Chen:2019a}. It allows to optimize the growth conditions and to boost the development of nanowire-based devices. It can be extended to a large variety of semiconductor materials, with a particular interest for polycrystalline layers, inhomogeneous materials, micro-crystals and nanostructures.

\section*{Acknowledgments}
This project has been supported by the French government in the frame of the "Programme d'Investissement d'Avenir" - ANR-IEED-002-01 and by the ANR projects Nanocell (ANR-15-CE05-0026) and Hetonan (ANR-15-CE05-0009).
This work was also partly funded in the frame of the EMPIR 19ENG05 NanoWires project. The EMPIR (European Metrology Programme for Innovation and Research) initiative is co-funded by the European Union's Horizon 2020 research and innovation programme and the EMPIR Participating States.

\appendix
\section{Room-temperature CL spectra for different injection levels}

\begin{figure*} 
	\includegraphics[width=0.8\textwidth]{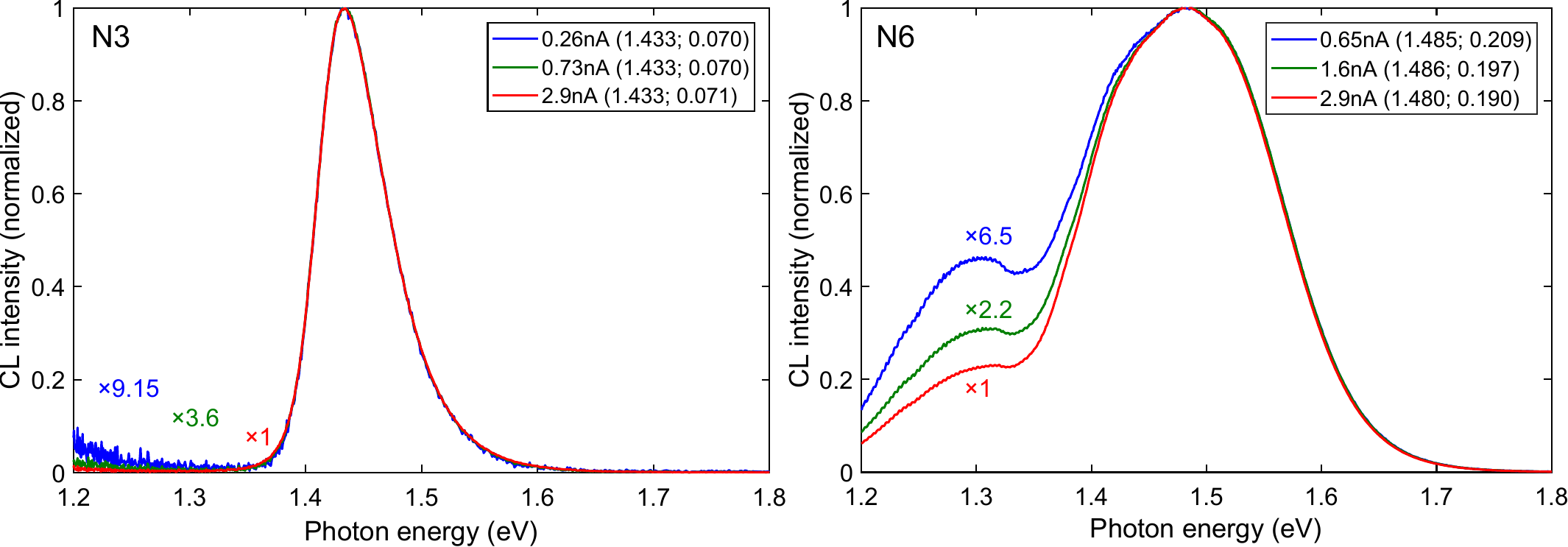}
	\caption{\label{fig:Injection}Room-temperature CL spectra of samples N3 and N6 for different excitation currents (inset: current of the electron beam, peak energy and FWHM in eV).}
\end{figure*}

Room-temperature CL spectra measured under different injection levels are plotted in Fig.~\ref{fig:Injection}. It shows that the excitation current has no impact on the position and width of the main emission peak. As expected, the low-energy defect peak is saturated and less visible for increasing injection currents (sample N6).

\section{Additional fits of room-temperature CL spectra}

In Figure~\ref{fig:ptype_RT}, room-temperature CL spectra measured on p-GaAs:Be thin films are plotted and compared to fits using the generalized Planck law with the fixed physical thickness (a,b) and a effective thickness fitted to in order to account for the low absorption of the simple parabolic absorption model (c,d). The fitted parameters are given in the main article for the fixed physical thickness, and in Table~\ref{tab:pdoping_fit} in the case of fitted effective thicknesses.

In Figure~\ref{fig:ntype_RT}, room-temperature CL spectra measured on n-GaAs:Si thin films are plotted and compared to fits using the generalized Planck's law with the fixed physical thickness (a,b) and a effective thickness fitted to in order to account for the low absorption of the simple parabolic absorption model (c,d). The fitted parameters are given in the main article for the fixed physical thickness, and in Table~\ref{tab:ndoping_fit} in the case of fitted effective thicknesses.

For the sake of simplicity and to limit the number of fitting parameters, our analysis in the main text makes use of fixed physical thicknesses. For both types of doping above $10^{18}$~cm$^{-3}$, the same bandgaps are found within about 1~meV, and the electron Fermi levels of n-doped samples are consistent with each other. However, for lower carrier concentrations, effective thicknesses lead to significantly lower BGN values.

\begin{table} 
	\caption{Optimal fit parameters obtained when the thickness is varied (bandgap $E_g$, Urbach tail $\gamma$, effective thickness $d_{\rm eff}$) for CL spectra measured on p-GaAs thin films at room temperature. Carrier densities from Hall effect measurements are indicated.}
	\label{tab:pdoping_fit}
	\begin{tabular}{cccccc}
		\hline\hline
		sample & carrier density & $E_g$ & $\gamma$ & $d_{\rm eff}$\\
		& cm$^{-3}$  & eV & meV & $\rm{\mu m}$\\
		\hline
		P1 & $2.7\times 10^{17}$ &  1.4214 & 8.8 & 6.20 \\ 
		P2 & $9.4\times 10^{17}$ &  1.4139 & 13.5 & 5.81 \\ 
		P3 & $2.4\times 10^{18}$ & 1.3977 & 17.0 & 2.99 \\ 
		P4 & $2.9\times 10^{18}$ & 1.3961 & 17.1 & 2.86 \\ 
		P5 & $1.0\times 10^{19}$ & 1.3898 & 17.9 & 1.51 \\ 
		P6 & $1.4\times 10^{19}$ & 1.3776 & 17.6 & 0.00 \\ 
		\hline\hline
	\end{tabular}
\end{table}

\begin{table}
	\caption{Optimal fit parameters obtained when the thickness is varied (bandgap $E_g$, electron Fermi level $E_{fc}$, Urbach tail $\gamma$, effective thickness $d_{\rm eff}$) for CL spectra measured on p-GaAs thin films at room temperature. Carrier densities from Hall effect measurements are indicated.}
	\label{tab:ndoping_fit}
	\begin{tabular}{cccccc}
		\hline\hline
		sample & carrier density & $E_g$ & $\gamma$ & $E_{fc}$ & $d_{\rm eff}$\\
		& cm$^{-3}$  & eV & meV & meV & $\rm{\mu m}$\\
		\hline
		N1 & $2.2\times10^{17}$ & 1.4157 & 7.1 & - & 3.09   \\ 
		N2 & $4.0\times10^{17}$ & 1.4171 & 7.5 & - & 3.31   \\ 
		N3 & $9.4\times10^{17}$ & 1.4078 & 12.9 & 17.7 & 0.23  \\ 
		N4 & $1.8\times10^{18}$ & 1.4043 & 16.1 & 47.6 & 0.88  \\ 
		N5 & $3.9\times10^{18}$ & 1.3841 & 20.9 & 112.8 & 0 \\ 
		N6 & $7.6\times10^{18}$ & 1.3772 & 23.1 & 173.1 & 0 \\ 
		\hline\hline
	\end{tabular}
\end{table}

\begin{figure*}
	\includegraphics[width=0.8\textwidth]{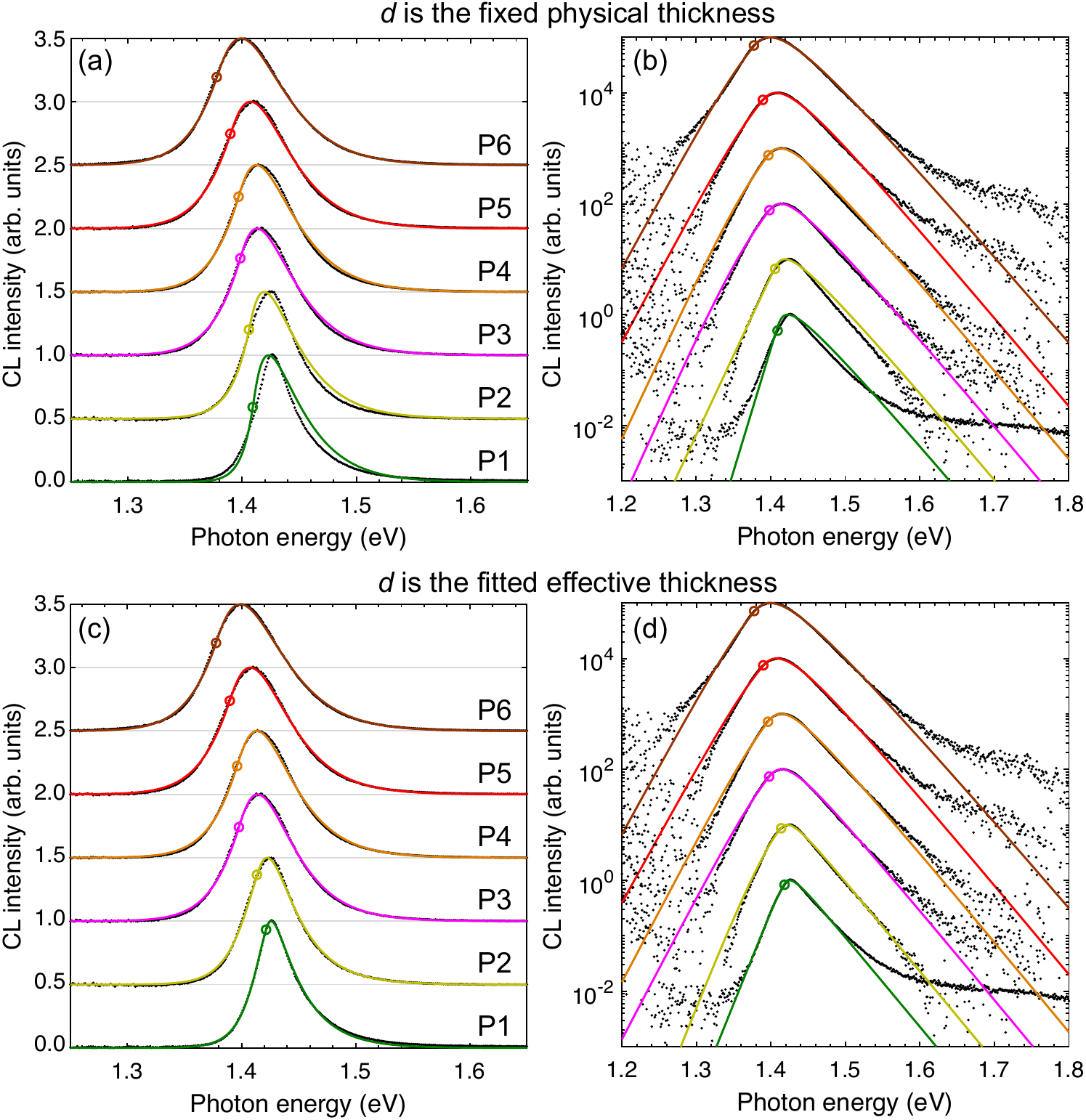}
	\caption{\label{fig:ptype_RT}CL spectra of p-GaAs:Be thin films measured at room temperature (dots) and the fits with the generalized Planck's law (lines) plotted in linear scale (a,c) and semi-log scale (b,d). Open circles superimposed on the CL spectra mark the bandgaps determined from the fits. (a,b) The physical thickness is used, and the fitting parameters are given in the main article, Table II. (c,d) An effective thickness is fitted in order to account for the low absorption of the simple parabolic absorption model, and the fitting parameters are given in Table~\ref{tab:pdoping_fit}.}
\end{figure*}

\begin{figure*}[h!]
	\includegraphics[width=0.8\textwidth]{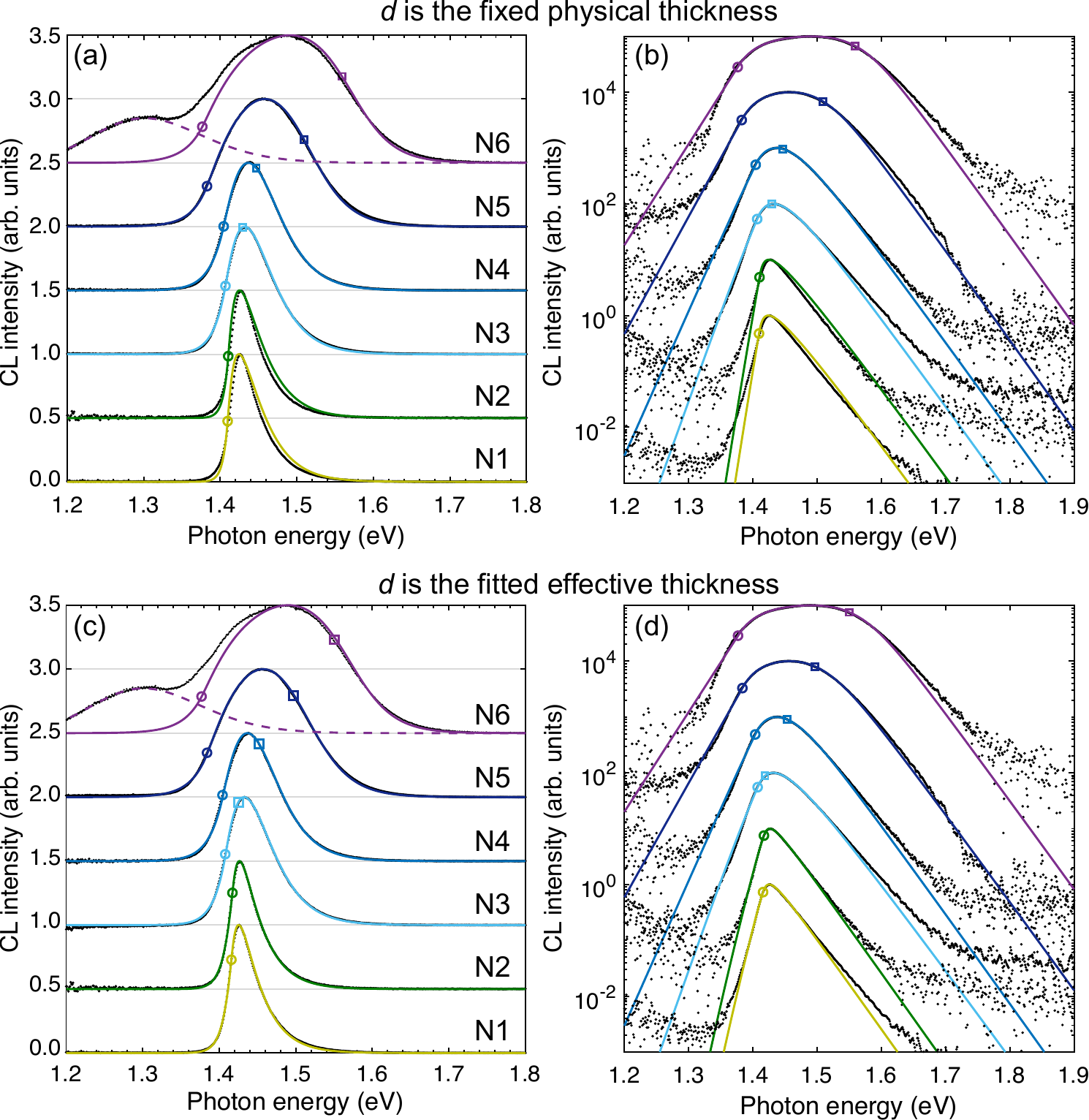}
	\caption{\label{fig:ntype_RT}CL spectra of n-GaAs:Si thin films measured at room temperature (dots) and the fits with the generalized Planck's law (lines) plotted in linear scale (a,c) and semi-log scale (b,d). Circles and squares mark the fitted bandgaps and electron Fermi levels, respectively. (a,b) The physical thickness is used, and the fitting parameters are given in the main article, Table III. (c,d) An effective thickness is fitted in order to account for the low absorption of the simple parabolic absorption model, and the fitting parameters are given in Table~\ref{tab:ndoping_fit}.}
\end{figure*}




\bibliography{CL_doping_thin_film}

\end{document}